\documentclass[prb,twocolumn,showpacs,floatfix,amsmath,amssymb,superscriptaddress]{revtex4-1}

\usepackage{amsfonts}
\usepackage{stmaryrd}
\usepackage{bbm}
\usepackage{mathrsfs}
\usepackage{tipa}
\usepackage{amssymb}
\usepackage{float}
\usepackage{txfonts}
\usepackage{graphicx}
\usepackage{dcolumn}
\usepackage{epstopdf}
\usepackage[colorlinks,linkcolor=blue,urlcolor=blue,citecolor=blue]{hyperref}
\usepackage{multirow}
\usepackage{subfigure}
\usepackage{url}
\usepackage{setspace}

\begin{document}

\title{Revealing the immediate formation of two-fold rotation symmetry in charge-density-wave state of Kagome superconductor CsV$_3$Sb$_5$ by optical polarization rotation measurement}

\author{Qiong Wu}
\affiliation{International Center for Quantum Materials, School of Physics, Peking University, Beijing 100871, China}

\author{Z. X. Wang}
\affiliation{International Center for Quantum Materials, School of Physics, Peking University, Beijing 100871, China}

\author{Q. M. Liu}
\affiliation{International Center for Quantum Materials, School of Physics, Peking University, Beijing 100871, China}

\author{R. S. Li}
\affiliation{International Center for Quantum Materials, School of Physics, Peking University, Beijing 100871, China}

\author{S. X. Xu}
\affiliation{International Center for Quantum Materials, School of Physics, Peking University, Beijing 100871, China}

\author{Q. W. Yin}
\affiliation{Department of Physics and Beijing Key Laboratory of Opto-electronic Functional Materials and Micro-nano Devices, Renmin University of China, Beijing 100872, China}

\author{C. S. Gong}
\affiliation{Department of Physics and Beijing Key Laboratory of Opto-electronic Functional Materials and Micro-nano Devices, Renmin University of China, Beijing 100872, China}

\author{Z. J. Tu}
\affiliation{Department of Physics and Beijing Key Laboratory of Opto-electronic Functional Materials and Micro-nano Devices, Renmin University of China, Beijing 100872, China}

\author{H. C. Lei}
\affiliation{Department of Physics and Beijing Key Laboratory of Opto-electronic Functional Materials and Micro-nano Devices, Renmin University of China, Beijing 100872, China}

\author{T. Dong}
\affiliation{International Center for Quantum Materials, School of Physics, Peking University, Beijing 100871, China}

\author{N. L. Wang}
\email{nlwang@pku.edu.cn}
\affiliation{International Center for Quantum Materials, School of Physics, Peking University, Beijing 100871, China}
\affiliation{Beijing Academy of Quantum Information Sciences, Beijing 100913, China}

\begin{abstract}

We report the observation of two-fold rotation symmetry in charge density wave (CDW) state in the newly discovered Kagome superconductor CsV$_3$Sb$_5$. Below its CDW transition temperature ($T_{CDW}$), the polarization rotation of the reflected laser beam promptly emerges and increases close to about 1 mrad, and the rotation angle shows two-fold rotation symmetry. With femtosecond laser pulse pumping, the rotation angle can be easily suppressed and then recovers in several picoseconds accompanied with coherent oscillations. Significantly, the oscillations in the signal also experience a 180 degree periodic change. Our investigation provides clear optical evidence for the formation of nematic order with two-fold rotation symmetry just below $T_{CDW}$. The results imply a immediate development of nematicity and possible time-reversal symmetry breaking in CDW state of CsV$_3$Sb$_5$.

\end{abstract}

\maketitle

Materials with Kagome structure often have very special physical properties, such as anomalous Hall effect\cite{Nakatsuji2015}, chiral edge state\cite{Yin2020}, and can also form spin liquid states\cite{Simeng2011,Han2012}. Recently, a new family of Kagome metal AV$_3$Sb$_5$ (A=K, Cs, Rb) has attracted tremendous attention due to its rich physical states and interactions \cite{Ortiz2019, Ortiz2020}. In those systems, vanadium atoms form a Kagome grid, and alkali metal atoms form a hexagonal structure. The AV$_3$Sb$_5$ family experiences a first-order structure transition and meanwhile forms a three-dimensional (3D) 2$\times$2$\times$2 charge density wave (CDW) order in the temperature range of 80-100 K \cite{PhysRevX.11.031026,Tan2021a,Li2021a,PhysRevB.104.165110}, in which the 2D Kagome lattice exhibits a 2$\times$2 structural distortion with three different V-V bond lengths. The 2$\times$2 CDW wave vector is equivalent to the momenta that connect the three M points in the hexagonal Brillouin zone, i.e. 3\textbf{Q} CDW \cite{Feng2021,KunJiangTaoWuJia}. Within the 2$\times$2 unit cell, the CDW structure contains the two trimers and one hexamer of V atoms, which is named as tri-hexagonal (TrH) structure or "inverse star of David". A $\pi$-phase shift between the adjacent TrH Kagome layer can lead to 3D 2$\times$2$\times$2 structural modulation  \cite{Tan2021a,Wang2021,KunJiangTaoWuJia,PhysRevMaterials.5.L111801}, being consistent with X-ray scattering measurement \cite{Miao2021,Tan2021a,Li2021a,KunJiangTaoWuJia}. Recent measurements also indicate presence of 2$\times$2$\times$4 superstructure, suggesting possible different interlayer stackings \cite{PhysRevX.11.041030,Wu2022,Zhang2022}. Further lowering of the temperature leads to the emergence of superconductivity at the critical temperature T$_c$ below 3 K, indicating the coexistence of CDW and superconductivity. Very interestingly, giant anomalous Hall effect (AHE) in AV$_3$Sb$_5$ was observed \cite{Yang2020,Yu2021a} in the absence of long-range magnetic order \cite{Kenney2020}.Besides, the scanning tunneling microscopy (STM) measurements \cite{Jiang2021,PhysRevB.104.075148,PhysRevB.104.035131} revealed that the three pairs of vector peaks corresponding to the 3\textbf{Q} CDW order in defect-free areas have different intensities, being attributed to the chiral CDW order. Theoretically, a chiral flux phase was proposed to explain the TRSB and AHE effect \cite{Feng2021,KunJiangTaoWuJia}. The proposed chiral flux phase and TRSB in CDW state were supported by recently zero-field muon-spin relaxation/rotation ($\mu$SR) measurements \cite{Mielke2022,yu2021evidence}. However, a very recent spin-polarized STM measurement found no indication of proposed chiral flux current phase or TRSB \cite{PhysRevB.105.045102}. AV$_3$Sb$_5$ (A = K, Cs, Rb) provides a new quantum platform to explore the intriguing interplay between topology, geometrical frustration and symmetry-breaking orders.

Identifying rotation symmetry breaking and/or TRSB in those compounds has become crucial topic in the current research on AV$_3$Sb$_5$ Kagome system as it is intimately related to the mechanisms for different orders including superconductivity \cite{KunJiangTaoWuJia}. It is known that the TRSB can lead to polar Kerr effect, that is, a linearly polarized light would have its polarization axis rotated upon reflection, in the meantime, the anisotropy in reflection allows to probe the rotation symmetry breaking \cite{Nandkishore2011}. Besides, a number of experimental measurements, including magnetoresistance \cite{Xiang2021a}, NMR \cite{Nie2022}, $\mu$SR \cite{yu2021evidence}, Raman \cite{Li2021,Wu2022} and coherent phonon spectroscopies \cite{PhysRevB.104.165110,PhysRevMaterials.5.L111801}, indicate that the CDW state is electronically nematic with only twofold (C$_2$) rotation symmetry at low temperature. Although the $\pi$-phase shift stacking between the neighboring layers can naturally explain the reduction of rotation symmetry from C$_6$ to C$_2$ in the CDW state, the onset of electronic nematicity was reported to be usually below 70 K, being well separated from CDW transition temperature  \cite{KunJiangTaoWuJia}.

Here we perform the static state and ultrafast time-resolved optical polarization rotation measurement to investigate the symmetry breakings in CsV$_3$Sb$_5$. When CsV$_3$Sb$_5$ cools down into CDW state, the polarization rotation of the reflected laser beam immediately emerges and reaches the value of about 1 mrad. During the whole CDW temperature range, the polarization rotation angle shows prominent polarization dependence on probe beam which includes a d-wave-like component. In the pump-probe investigation, we observed the coherent phonon-induced oscillations in the recovery process of transient polarization rotation. The oscillations also keep the same two-fold rotation symmetry with static result and undergoes a $\pi$ phase shift at four specific polarization orientations. With pump fluence increasing, the pump-induced diffrential rotation and oscillations can be easily suppressed. Our work reveals development of the electronic nematic order with a C$_2$ rotation symmetry just below T$_{CDW}$, offering new insights in understanding and controlling the novel properties of this system.

\begin{figure*}[t]
	\centering
	\includegraphics[width=17cm]{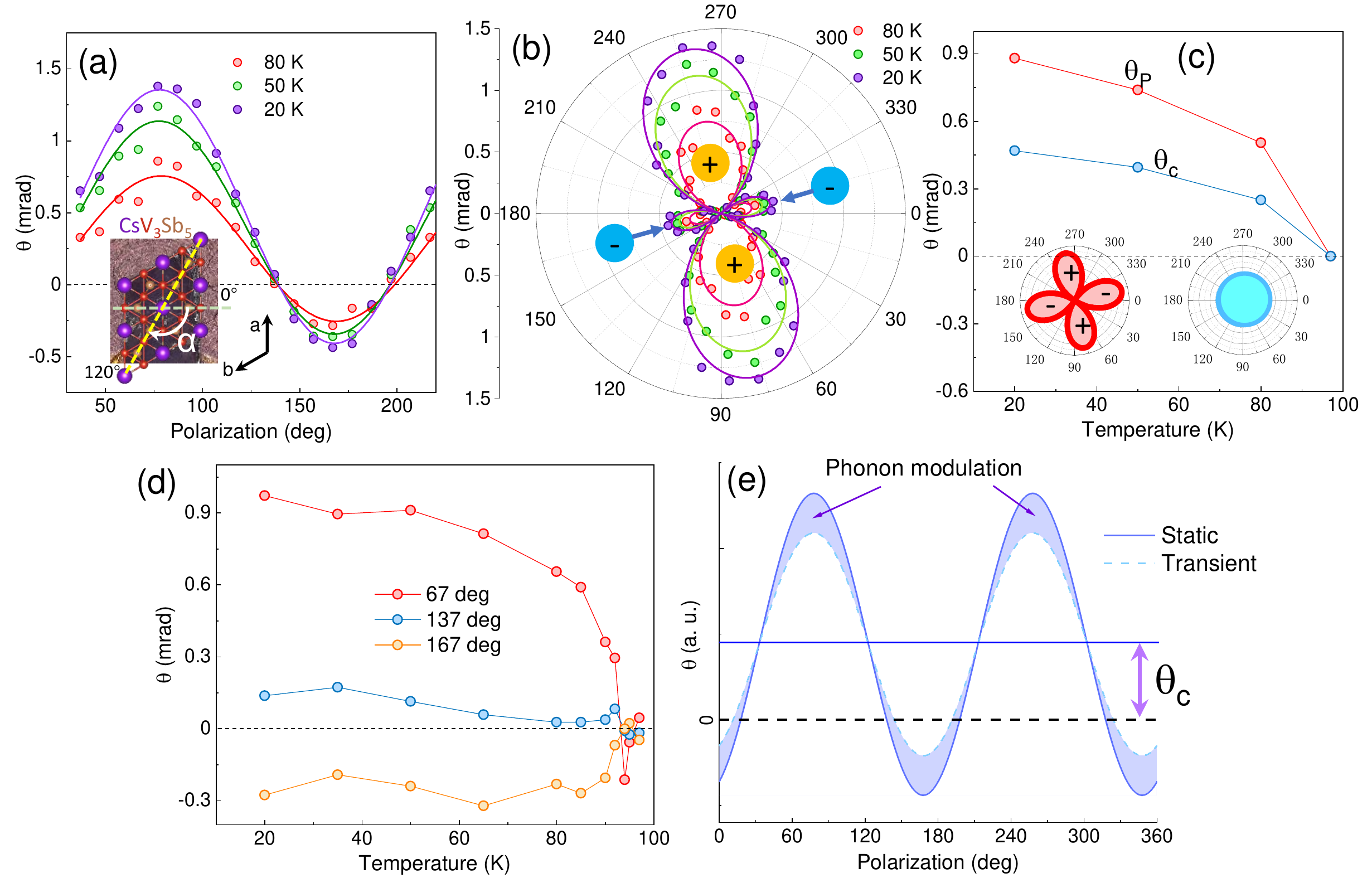}\\
	\caption{\textbf{The static state polarization rotation for CsV$_3$Sb$_5$.} (a) The $\theta$ as a function of incident polarization at 80 K, 50 K, and 20 K. Note that $\theta$ have two signes at different polarizations. The solid curves are cosine fitting results. The inset illustrates the recognized crystal orientation according to the shape of edge and labeled polarization in experiment. (b) The polar coordinates plot of panel (a). (c) Fitted parameters of $\theta_p$ and $\theta_c$ in (a,b) from Eq. (1). The left (right) inset sketches the polarization dependent (independent) components. (d) $\theta$ in the temperature range of 5-100 K along 3 different polarizations. (e) Schematic of static (solid curve) and transient (dashed curve) $\theta$ modulated by coherent phonons.}\label{Fig:1}
\end{figure*}

Single crystals of CsV$_3$Sb$_5$ were grown from Cs ingot (purity 99.9\%), V powder (purity 99.9\%) and Sb grains (purity 99.999\%) using the self-flux method, which is similar to the growth of RbV$_3$Sb$_5$\cite{Yin2021}. For static reflection experiment, we use the Ti:sapphire oscillator with 800 nm center wavelength, 80 MHz repetition rate, and 100 fs pulse duration to measure polarization rotation $\theta$. The probe beam nearly-normally illuminates the sample and the reflected probe beam was divided into two beams with vertical and horizontal polarization by a beamsplitter cube, respectively. The two beams were collected respectively by two detectors and the corresponding differential signal can be obtained (for more details, see Supplemental Information). In pump-probe investigation, we utilize the amplified Ti:sapphire laser with 800 nm center wavelength, 1 kHz repetition rate, and 85 fs pulse duration to carry out the transient dichrism experiment. During the polarization- and temperature- dependent pump-probe experiments, the pump fluence was set to be 14 $\mu$J cm$^{-2}$.

Figure \ref{Fig:1}(a) and 1(b) illustrate the polarization orientation dependence of rotation angle $\theta$ at several typical temperatures below $T_{CDW}$. The $\theta$ forms a surprisingly petaiod pattern in 0-360 degree, which unambiguously reveals a two-fold rotation symmetry. It can be seen that the polarization orientation ($\alpha$) dependence of $\theta$ at these temperatures share the same rotation symmetry. By using cosine function:
\begin{equation}
\begin{aligned}
    \theta(\alpha) = \theta_p \cdot cos(2\pi\alpha/P-\beta)+\theta_c\\
\end{aligned}
\end{equation}
the $\theta$ result in Fig. \ref{Fig:1}(a) and (b) can be well fitted. Where the $\theta_p$ ($\theta_c$) represents the amplitude of polarization dependent (independent) rotation component, which will be discussed later; $P$ represents the rotation period with the value of 180 deg; $\beta$ represents the polarization at which the $\theta$ reaches maximum, which remains about 75 and 255 deg in our experiment.

\begin{figure*}[t]
	\centering
	\includegraphics[width=18cm]{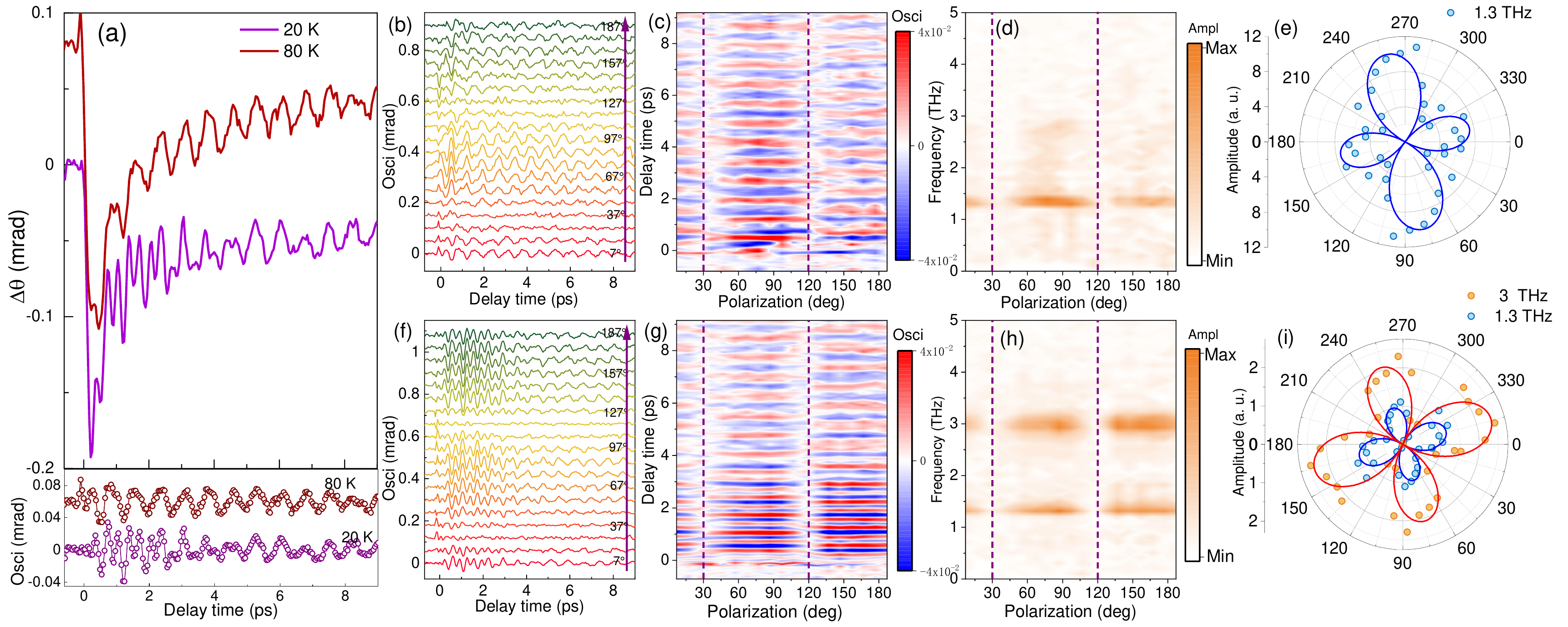}\\
	\caption{\textbf{Polarization-dependent coherent oscillations in pump-probe measurement for CsV$_3$Sb$_5$.}
	(a) Pump-probe rotation dynamics at 20 K and 80 K. Lower panel: extracted phonon oscillations.
	(b,f) The oscillations in optical pump-probe signals at different polarizations. (c,g) Two-dimensinal plot of oscillation as a function of both delay time and $\alpha$. (d,h) Fast Fourier transform results with delay time. (e,i) $\alpha$-dependent of integrated strength of phonon oscillations from (d,h). The result in panels (b)-(e) are obtained at 80 K, where that in panels (f)-(i) are obtained at 20 K.
	}\label{Fig:2}
\end{figure*}

The fitting parameters of $\theta_p$ and $\theta_c$ are shown in Fig. \ref{Fig:1}(c). Note that there are two terms in Eq. (1), which are schematically plotted in the inset of Fig. \ref{Fig:1}(c) where $\theta_p \cdot cos(2\pi\alpha/P-\beta)$ term is d-wave like and $\theta_c$ term is a constant. In the whole CDW state, $\theta_p$ is much larger than $\theta_c$. Figure \ref{Fig:1}(d) presents the temperature-dependent static polarization rotation of CsV$_3$Sb$_5$ along three typical polarizations. In normal state, $\theta$ keeps to be zero. When temperature get reaches 94 K, $\theta$ immediately raises. At $\alpha$ = 67 degree, $\theta$ remains a positive value. At 20 K, the rotation angle reaches 1 mrad, such a large $\theta$ can be compared with that of some magnetic materials\cite{Di1994} ; At $\alpha$ = 137 degree, $\theta$ is nearly independent with temperature above 60 K and slightly increases at lower temperature; At $\alpha$ = 167 degree, the $\theta$ turns to be negative below $T_{CDW}$. Below 94 K, there is no other unique temperature point, indicating the development of two-fold rotation symmetry just below $T_{CDW}$. The signal with identical properties has also been repeated in another CsV$_3$Sb$_5$ sample (see Supplemental Information). The phonon modulation on $\theta$ is illustrated in Fig. \ref{Fig:1}(e), which will be discussed later. It deserves to remark that, unlike other experimental probes which indicated a nematic charge order only at low temperature\cite{Xiang2021a,Nie2022,yu2021evidence,Li2021,Wu2022}, the polarization rotation appears to be extremely sensitive to the rotation symmetry breaking from C$_6$ to C$_2$. The rotation of polarization and associated twofold symmetry of reflected laser beam develop immediately as long as the CDW phase transition is established. This observation is consistent with the measurement of anormalous Hall effect with the signal appearing just below T$_{CDW}$.

After investigating the static state optical polarization rotation, we carry out the optical pump-polarization rotation probe experiment. Different from other CDW materials, the CDW state of CsV$_3$Sb$_5$ can be affected by optical pulse, thus the CDW order-induced rotation can also be modulated. After femtosecond optical pulse pumping, the $\theta$ instantaneously drops with 0.2 mrad amplitude, and relaxes within several picoseconds, as shown in Fig. \ref{Fig:2} (a). After subtracting the exponential decay process, the coherent oscillations can be extracted. At 80 K, there is only one oscillation mode (f = 1.3 THz), while two oscillation modes (f=1.3 and 3.0 THz, respectively) appearing at 20 K. These two modes have been observed by optical pump-probe spectroscopy \cite{PhysRevMaterials.5.L111801,PhysRevB.104.165110}, where the 1.3 THz phonon involve the CDW structure modulation and the 3 THz phonon was likely to be linked to the uniaxial order.

By rotating the incident polarization of probe beam, we can measure the coherent phonon oscillations in dynamics at different crystal orientations. At 80 K, when changing $\alpha$ from 7 to 27 deg, overall oscillations become weak (Fig. \ref{Fig:2} (b)). At 37 deg, there are only two or three periods that can be observed. With $\alpha$ further increasing, the phonon experiences a $\pi$ phase difference whose amplitude enhances again. The same phenomenon also appears around 120 deg. We plot the colormap $\Delta\theta$ as a function of both polarization
direction and delay time in Fig. \ref{Fig:2} (c). Two nodes at 30 and 120 deg can be distinguished. After carrying out the fast Fourior transform (FFT), it can be seen that only 1.3 THz phonon mode is included (Fig. \ref{Fig:2} (d)). The phonon strength can be well fitted by a cosine function (Fig. \ref{Fig:2}(e)). At 20 K, there are two phonon modes, which both disappear at around 30 and 120 deg (Fig. \ref{Fig:2}(f-i)). And the polarization dependence of their strength also obeys the cosine function. In contrast, the coherent oscillations measured in transient reflectivity measurement do not have the $\pi$ phase shift during the polarization change (see Supplemental Information), thus we can ensure that the $C_2$ rotation symmetry of coherent oscillation solely comes from the polarization rotation dynamics in CsV$_3$Sb$_5$.

\begin{figure*}[t]
 \centering
 \centering\includegraphics[width=17cm]{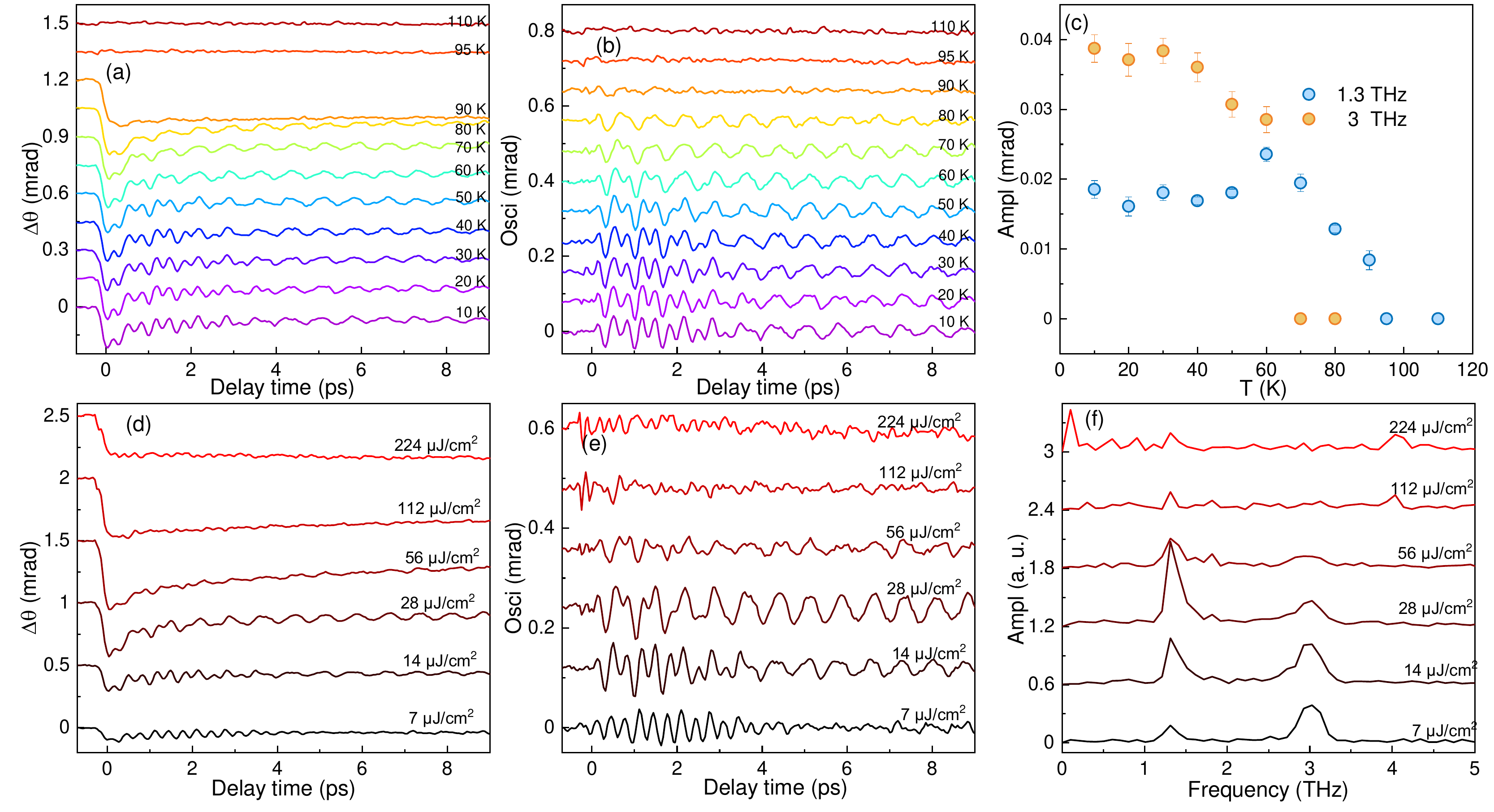}\\
 \caption{
\textbf{Temperature- and fluence-dependent transient polarization rotation signal of CsV$_3$Sb$_5$.} (a) Photo-induced $\Delta$$\theta$ as a function of time delay in temperature range from 10 to 110 K (b) Extracted oscillations from panel (a). (c) The amplitudes of two phonon modes. (d) The transient polarization rotation dynamics under different pump fluences. (e) Extracted oscillations from panel (d). (f) FFT result of panel (e). The polarization are fixed at $\alpha$ = 167 deg. And the results in (d)-(f) are abtained at 20 K. Data are offset for clarity} \label{Fig:3}
\end{figure*}

The temperature variation can dramatically influence the dynamics. To monitor the effect clearly, we fix the polarization at 167 deg and carry out the temperature dependent experiment. As shown in Fig. \ref{Fig:3}(a), below 80 K, the dynamics roughly includes a fast and a slow decay processes, and on top of them, the coherent oscillations are prominent. At 90 K, there is only a slow decay process survived, and the oscillations are almost invisible. Above 95 K, the dynamics and oscillations completely vanish. This phenomenon verifies that the polarization rotation dynamics comes from the CDW state. The extracted oscillations are illustrated in Fig. \ref{Fig:3}(b). After fitting the time-domain oscillations, the strength of two phonon modes are depicted in Fig. \ref{Fig:3}(c). At low temperature, the 3 THz phonon is about two times stronger than 1.3 THz phonon. With temperature increasing, the 3 THz phonon weakens while 1.3 THz enhances, representing a competing relation. Above 60 K, the 3 THz phonon mode disappears and 1.3 THz phonon turns to be suppressed. This phenomenon is similar to that observed in the optical reflectivity pump-probe measurement \cite{PhysRevB.104.165110}.

We also study the influence of pump fluence to the polarization rotation dynamics at 20 K shown in Fig. \ref{Fig:3}(d-f). Below 56 $\mu$J cm$^{-2}$, the dynamics has the positve correlation with  pump fluence accompanied by degeneration of coherent phonons (Fig. \ref{Fig:3}(d)). Under high pump fluence ($\geq$ 56 $\mu$J cm$^{-2}$), the dynamics is suppressed. It is in line with the previous reports that CDW state in CsV$_3$Sb$_5$ can be restrained by optical illumination \cite{PhysRevB.104.165110}. In addition, at 56 $\mu$J cm$^{-2}$, the $\Delta$ $\theta$ drops to be -0.6 mrad, which is comparable to the static $\theta$ value. It is reasonable to consider that at this pump fluence, the symmetry breaking is remarkably suppressed. At the lowest pump fluence, the oscillation damps within 5 picoseconds (Fig. \ref{Fig:3}(e)). Increasing to 14 $\mu$J cm$^{-2}$, the 1.3 THz phonon mode with long lifetime emerges. At 28 $\mu$J cm$^{-2}$, the 1.3 THz phonon oscillation becomes dominated. With fluence further increasing, both phonon modes become weaker. The FFT result (Fig. \ref{Fig:3}(f)) emphasizes this transition and additionally shows the appearance of 4.1 THz phonon.

There are two possibilities to explain the observed polarization rotation of reflected laser beam. One is the proposed TRSB below $T_{CDW}$. Theoretically, it is suggested that the lattice distortion with inverse star of David pattern leads to the emergence of a chiral flux phase with two separate flux loops formed respectively on trimers and hexamer of V lattices. This chiral flux phase breaks the time-reversal symmetry and naturally explains the 3\textbf{Q} CDW order and giant AHE \cite{Feng2021}. It is the most exotic and appealing state in those compounds, which is reminiscent of chiral flux formation proposed to produce the quantum anomalous Hall effect in a Honeycomb lattice\cite{PhysRevLett.61.2015} and loop-current phase in cuprate high temperature superconductors \cite{PhysRevB.55.14554}. Searching for the TRSB phase has become a crucial topic in the current resarch on AV$_3$Sb$_5$. Two recent zero-field muon-spin relaxation/rotation ($\mu$SR) measurements have yielded evidence for the TRSB \cite{Mielke2022,yu2021evidence}. However, the TRSB signal by $\mu$SR measurement on CsV$_3$Sb$_5$ develops only below 70 K, which is significantly lower than the CDW transition temperature $T_{CDW}$=94 K \cite{yu2021evidence}. The present measurement result seems to be consistent with the scenario.

Another possibility is that the observed optical polarization rotation is arising from the birefringence effect which may be related to the in-plane anisotropy caused by the $\pi$-phase shift interlayer stacking. Nematic order has been observed in many high-temperature superconductors and can cause d-wave-like pattern of optical response. Birefringence is one of the well-known optical property of anisotropic materials and is widely used in fabricating quarter and half wave plates and optical devices. It is also possible that both TRSB and birefringence contribute to the rotation of polarization. For AV$_3$Sb$_5$ below $T_{CDW}$, the formation of inverse-David-star lattice distortion within a single layer does not break the C$_6$ rotation symmetry. The staggerly stacked inverse-David-star layers of 3D CDW break the C$_6$ rotation symmetry and lead to C$_2$ rotation symmetry or nematic electron state. Thus, one may tend to attribute the constant and two-fold rotational components of measured polarization rotation to the Kerr signal arising from TRSB and the birefringence effect caused by the anisotropy of interlayer stackings, respectively. Further explorations are still needed.

On the other hand, certain anomaly really develops near 60 K which has been observed in many different experimental probes. In the present measurement, the coherent phonon oscillations also show abnormal feature near 60 K, similar to the observations in the reflectivity pump-probe measurement \cite{PhysRevB.104.165110,PhysRevMaterials.5.L111801}. The invariant symmetry of oscillations reveals the single lattice symmetry below T$_{CDW}$. Note that the observed phonons possess A$_1g$ symmetry. We emphasize that the displacement of atoms arising from coherent phonon modes changes the equilibrium position of V or other atoms and then influence the CDW strength. The CDW order has strong correlation with the distorted lattice structure and chrality, thus the coherent oscillations of phonons could modulate $\theta$. We note that the $\alpha$ dependence of coherent phonon amplitude in Fig. \ref{Fig:2} (e) and (i) is similar to the $\alpha$ dependent term $\theta_p \cdot cos(2\pi\alpha/P-\beta)$, showing in the left inset of Fig. \ref{Fig:1}(c). It demonstrates that phonon oscillations only modulate the anisotropic $\theta_p$ component. As shown in Fig. \ref{Fig:1}(f), the modulation caused by coherent phonons can lead to two nodes at $\alpha$=30 deg and 120 deg in the range of 0 - 180 deg with a $\pi$-phase shift in the phonon oscillation at the two sides. This explains the $\pi$-phase shift of coherent phonon oscillations and the intensity change observed in Fig. \ref{Fig:2}(c) and (g). Since there is no indication of symmetry breaking around 60 K in static and pump-probe experiments, the results point out that the observed anomaly below 60 K is less relevant with the two-fold rotation symmetry, which is likely due to the reorientation of stacked inverse star of David layers. 

In summary, optical polarization rotation is an effective method to investigate the detailed properties of symmetry breaking in a material. Our measurement reveals a unique phase transition property for CDW state in CsV$_3$Sb$_5$. Except for the charge oder formation below $T_{CDW}$, the remarkable bulk symmetry breaking state with two-fold rotation symmetry is simultaneously established.  Remarkably, the CDW-related coherent A$_1g$ phonons can modulate the strength of in-plane anisotropy, particularly the polarization dependent term caused by the stacking of Kagome layers, leading to a 180 deg periodic change in the oscillations specifically on the pump-probe signal. The observed polarization rotation can be contributed either from TRSB or birefringence effect caused by the anisotropy of interlayer stackings. Our measurement reveals that the anomaly around 60 K is less relevant with the bulk two-fold rotation symmetry breaking, but could be due to the reorientation of stacked inverse star of David layers. Furthermore, we observe that pump pulse can transiently suppress the CDW state. Our work offers new insights in understanding and controlling the novel properties of CsV$_3$Sb$_5$.

\begin{center}
\small{\textbf{ACKNOWLEDGMENTS}}
\end{center}

We thank the helpful discussion with Professor L. Y. Yang and Doctor L. Y. Liu. This work was supported by the National Key Research and Development Program of China (No. 2017YFA0302904, 2018YFE0202600), the National Natural Science Foundation of China (No. 11888101, 11822412 and 11774423), and Beijing Natural Science Foundation (Grant No. Z200005)

\bibliography{CsVSb-1}

\end{document}